\documentclass[aps,prl,twocolumn,showpacs,byrevtex]{revtex4}

\usepackage{times,mathptm}
\usepackage[dvips]{graphicx,color}
\usepackage{titlesec}
\newcommand{\bqa}{\begin{eqnarray*}}
\newcommand{\eqa}{\end{eqnarray*}}
\makeatletter
\renewcommand{\@seccntformat}[1]{}
\makeatother

\begin{document}

\title{Stabilization mechanism of clathrate H cages in a room-temperature superconductor LaH$_{10}$}

\author{Seho Yi, Chongze Wang, Hyunsoo Jeon, and Jun-Hyung Cho$^{*}$}
\affiliation{Department of Physics, Research Institute for Natural Science, and HYU-HPSTAR-CIS High Pressure Research Center, Hanyang
University, 222 Wangsimni-ro, Seongdong-Ku, Seoul 04763, Republic of Korea}
\date{\today}

\begin{abstract}
Lanthanum hydride LaH$_{10}$ with a sodalitelike clathrate structure was experimentally realized to exhibit a room-temperature superconductivity under megabar pressures. Based on first-principles calculations, we reveal that the metal framework of La atoms has the excess electrons at interstitial regions. Such anionic electrons are easily captured to form a stable clathrate structure of H cages. We thus propose that the charge transfer from La to H atoms is mostly driven by the electride property of the La framework. Further, the interaction between La atoms and H cages induces a delocalization of La-5$p$ semicore states to hybridize with H-1$s$ state. Consequently, the bonding nature between La atoms and H cages is characterized as a mixture of ionic and covalent. Our findings demonstrate that anionic and semicore electrons play important roles in stabilizing clathrate H cages in LaH$_{10}$, which can be broadly applicable to other high-pressure rare-earth hydrides with clathrate structures.
\end{abstract}

\maketitle

\subsection{INTRODUCTION}
Metal hydrides have attracted much attention theoretically~\cite{CaH6-PANS2012, H3S-Theory, rare-earth-hydride-PRL2017, rare-earth-hydride-PANS2017, LaH10-PRB2019-Liangliang, LaH10-PRB2019-Chongze,YH10-Boeri} and experimentally~\cite{H3S-expt, ExpLaH10-PRL2019, ExpLaH10-Nature2019, ExpCeH9-Nat.Commun2019T.Cui, ExpCeH9-Nat.Commun2019-J.F.Lin, ExpYH6-arXiv2019} because of their promising possibility for the realization of room-temperature superconductivity (SC)~\cite{review-Zurek,review-Eremets}. Recently, first-principles density-functional theory (DFT) calculations and the Migdal-Eliashberg formalism predicted that the rare-earth metal hydrides such as yttrium hydride YH$_{10}$ and lanthanum hydride LaH$_{10}$ exhibit a room-temperature SC at megabar pressures~\cite{rare-earth-hydride-PRL2017, rare-earth-hydride-PANS2017}, which is driven by the quantum condensate of electron pairs, so-called Cooper pairs~\cite{BCS}. Motivated by these theoretical predictions, the conventional phonon-mediated SC of LaH$_{10}$ was experimentally observed with a superconducting transition temperature $T_{\rm c}$ of 250$-$260 K at a pressure of ${\sim}$170 GPa~\cite{ExpLaH10-PRL2019, ExpLaH10-Nature2019}. This $T_{\rm c}$ record of compressed LaH$_{10}$ is much higher compared to those of unconventional superconducting materials such as cuprates~\cite{Cuprate-Nature1993, Cuprate-Rev-Nature2010} and Fe-based superconductors~\cite{Fe-based-Zhao2008, Fe-based-Chen2009}. Therefore, the realization of room-temperature SC in rare-earth metal hydrides has launched a new era of high-$T_{\rm c}$ superconductors.

The superhydride LaH$_{10}$ has a hydrogen sodalitelike clathrate structure with the high crystalline symmetry of space group $Fm$$\overline{3}m$ (No. 225). As shown in Fig. 1a, this cubic structure is constituted by the fcc lattice of La atoms, where each La atom is surrounded by the cage of 32 H atoms. Hereafter, the La framework without H atoms is designated as LaH$_0$ (see Fig. 1b), whereas the H$_{32}$ cages without La atoms are designated as La$_0$H$_{10}$ (Fig. 1c). Since phonon-mediated electron pairing is responsible for the room-temperature SC of LaH$_{10}$~\cite{rare-earth-hydride-PRL2017, rare-earth-hydride-PANS2017, LaH10-PRB2019-Liangliang, LaH10-PRB2019-Chongze, ExpLaH10-Nature2019}, the identification of bonding nature is the beginning step towards a proper understanding of electron-phonon coupling. According to an earlier DFT calculation of LaH$_{10}$~\cite{rare-earth-hydride-PRL2017}, the bonding between La atoms and H$_{32}$ cages was assigned to purely ionic because of the lack of charge localization between the two constituents, while the H$-$H bonding in the H$_{32}$ cage was described by a weakly covalent character. Moreover, the Mulliken population analysis showed a significant charge transfer from La to H atoms, which was interpreted in terms of the lower electronegativity of La atom compared to that of H atom~\cite{rare-earth-hydride-PRL2017}. Meanwhile, the calculated partial electronic density of states (DOS) revealed a strong hybridization of La 4$f$ and H 1$s$ states near the Fermi energy $E_{\rm F}$~\cite{LaH10-PRB2019-Liangliang}, reflecting a covalent-like bonding character between La and H atoms. Therefore, it remains to be resolved whether the bonding nature between La atoms and H$_{32}$ cages is ionic or covalent.

\begin{figure}[ht]
\centering{ \includegraphics[width=8.0cm]{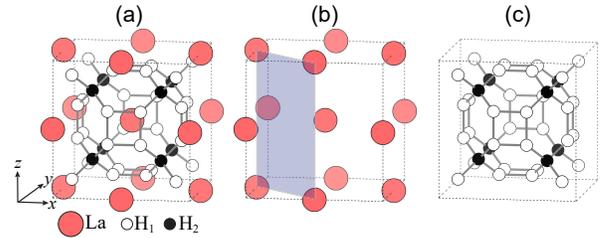} }
\caption{(a) Optimized structure of LaH$_{10}$ at 300 GPa, showing the H$_{32}$ cage surrounding a La atom. There are two different types of H atoms, H$_{1}$ and H$_{2}$, depending on the surrounding bond directions. The separated La framework (LaH$_{0}$) and H$_{32}$ cages (La$_{0}$H$_{10}$) are displayed in (b) and (c), respectively. The $x$, $y$, and $z$ axes point along the [001], [010], and [001] directions, respectively. A (220) plane is drawn in (b).}
\end{figure}

\begin{figure*}[ht]
\centering{ \includegraphics[width=16.0cm]{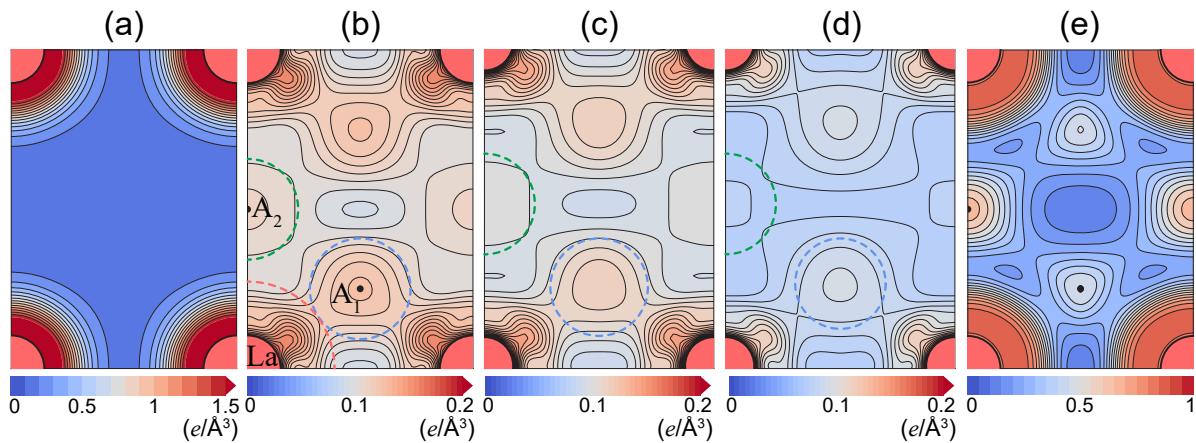} }
\caption{Calculated total charge density of LaH$_0$ at 300 GPa (a) with including semicore electrons and (b) without including semicore electrons. For comparison, the total charge densities of LaH$_0$ at 220 GPa and ambient pressure, obtained without including semicore electrons, are displayed in (c) and (d), respectively. The ELF of LaH$_0$, calculated at 300 GPa, is displayed in (e). In (b), A$_1$ and A$_2$ indicate the two anions in interstitial regions (see Fig. S2 in the Supplementary Information), while the dashed circles represent the muffin-tin spheres around La and A$_1$ (A$_2$) with the radii of 1.31 and 0.75 (0.75) {\AA}, respectively. The charge densities are drawn on the (220) plane with a contour spacing of 0.01 $e$/{\AA}$^3$ in (a) and a contour spacing of 0.005 $e$/{\AA}$^3$ in (b), (c), and (d). The ELF in (e) is drawn with a contour spacing of 0.05.}
\end{figure*}

In this paper, we perform the first-principles DFT calculations for LaH$_0$, La$_0$H$_{10}$, and LaH$_{10}$ to systematically analyze their charge densities and electronic structures. It is revealed that the La framework LaH$_0$ behaves as an electride under high pressure, where some electrons detached from La atoms are well localized in interstitial regions. Such loosely bound anionic electrons can be easily captured to form the clathrate H$_{32}$ cages composed of two H species (termed H$_1$ and H$_2$ in Fig. 1a). Therefore, donating electrons to the H$_{32}$ cages in LaH$_{10}$ are mostly attributed to anionic electrons created in LaH$_0$, rather than due to a charge transfer induced by electronegativity differences between La and H atoms. We also reveal that the interaction between the LaH$_0$ and La$_0$H$_{10}$ constituents gives rise not only to a hybridization of La 4$f$ and H$_1$ 1$s$ states around $E_{\rm F}$ but, more importantly, to a delocalization of La 5$p$ semicore states to hybridize with H$_1$ 1$s$ state. Based on these results, it is most likely that the bonding nature between La atoms and H$_{32}$ cages is characterized as a mixture of ionic and covalent. The present findings not only demonstrate that anionic and semicore electrons are of vital importance to yield the nature of the intricate bonding in LaH$_{10}$, but also have important implications for understanding the bonding and electronic properties of other rare-earth hydrides with sodalitelike clathrate structures.~\cite{rare-earth-hydride-PRL2017, rare-earth-hydride-PANS2017,YH10-Boeri,ExpCeH9-Nat.Commun2019T.Cui, ExpCeH9-Nat.Commun2019-J.F.Lin, ExpYH6-arXiv2019}

\begin{figure}[h!t]
\centering{ \includegraphics[width=8.0cm]{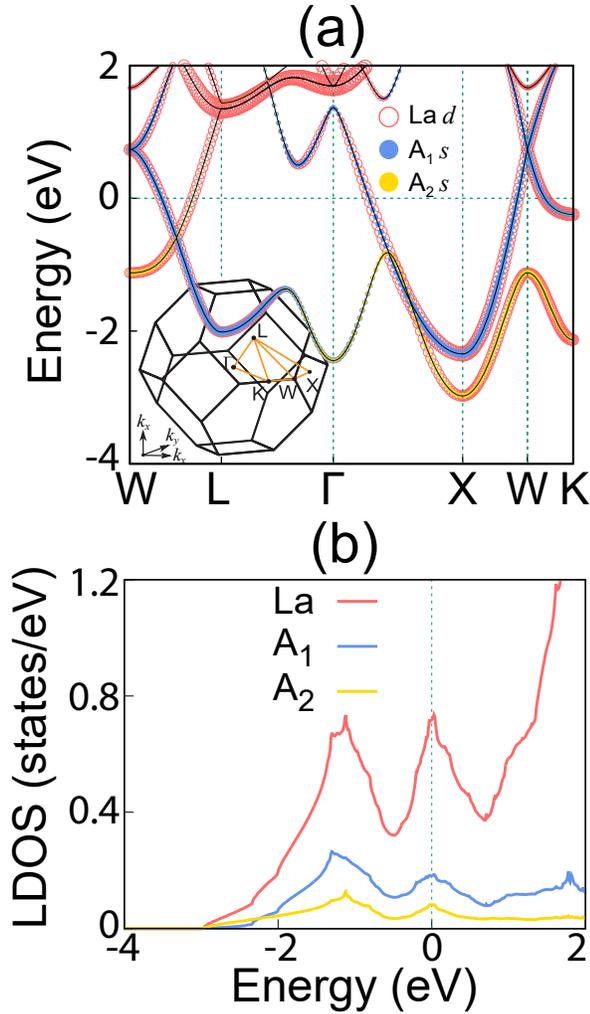} }
\caption{Calculated (a) band structure and (b) LDOS of LaH$_0$ at 300 GPa. In (a), the projected bands onto the La 5$d$, A$_1$, and $A_2$ $s$-like orbitals are represented by circles whose radii are proportional to the weights of the corresponding orbitals. The inset of (a) shows the Brillouin zone of the fcc primitive unit cell. In (b), the LDOS curves of La, A$_1$, and A$_2$ (pseudo)atoms are given. The energy zero represents $E_{\rm F}$.}
\end{figure}

\subsection{RESULTS}
We first present the total charge density of LaH$_0$, obtained using the DFT calculation (see the section of Methods). Here, the structure of LaH$_0$ (see Fig. 1b) is taken from the optimized structure of LaH$_{10}$ at a pressure of 300 GPa, where the lattice parameters of the conventional cubic unit cell are $a$ = $b$ = $c$ = 4.748 {\AA}. Figure 2a displays the calculated total charge density of LaH$_0$. It is seen that LaH$_0$ apparently exhibits a spherically symmetric charge distribution around each La atom, most of which is contributed from the 5$s^2$5$p^6$ semicore electrons. We note that the 5$s^2$ and 5$p^6$ semicore states are located at around $-$32 eV and $-$17 eV below $E_{\rm F}$ (see Fig. S1 in the Supplementary Information), respectively, which are well separated from the 5$d^1$6$s^2$ valence electrons. In order to reveal the charge distribution of valence electrons in LaH$_0$, we exclude the 5$s^2$5$p^6$ semicore electrons to plot the valence charge density in Fig. 2b. Interestingly, we find that some electrons detached from La atoms are well localized in the interstitial regions around the A$_1$ and A$_2$ sites (see Fig. 2b) on the (220) plane. The numbers of electrons within the muffin-tin spheres around the La, A$_1$, and A$_2$ (pseudo)atoms are 1.17$e$, 0.21$e$, and 0.19$e$, respectively. Here, each muffin-tin sphere is drawn with a dashed circle in Fig. 2b. It is thus likely that LaH$_0$ can be considered as an electride with a [La]$^{1.83+}$:1.83$e^{-}$ configuration. We note that the numbers of anionic electrons in the A$_1$ and A$_2$ pseudoatoms (or anions) are reduced with decreasing pressure (see Figs. 2c and 2d): i.e., A$_1$ (A$_2$) has $-$0.19$e$ ($-$0.17$e$) at 220 GPa and $-$0.15$e$ ($-$0.13$e$) at ambient pressure. As shown in Fig. 2e, the confinement of anionic electrons around the A$_1$ and A$_2$ sites is well confirmed at 300 GPa by the electron localization function (ELF)~\cite{ELF}, which is effective for the characterization of interstitial electrons in electride materials~\cite{ELF-elect1,ELF-elect2,ELF-elect3}. At high pressure, the overlap of atomic valence electrons increases their Coulomb repulsion, which in turn leads to the localization of anionic electrons in interstitial regions~\cite{hoffman,hosono,Li6P}. Thus, we can say that the electride characteristics of LaH$_0$ with the A$_1$ and A$_2$ anions become dominant as pressure increases.

To explore the peculiar feature of valence electronic states with anionic electrons, we present the band structure of LaH$_0$ in Fig. 3a. There exist two partially occupied bands crossing $E_{\rm F}$, which originate dominantly from the La 5$d$ orbitals and the A$_1$/A$_2$ $s$-like orbitals. Here, the band projections onto other orbitals are relatively minor: see Fig. S3 in the Supplementary Information. It is noticeable that the two bands show a strong hybridization of the La 5$d$ and A$_1$ (or A$_2$) $s$-like orbitals depending on the symmetry lines. Fig. 3(b) shows the local DOS (LDOS) within the muffin-tin spheres around the La, A$_1$, and A$_2$ (pseudo)atoms. We find that the LDOS for the A$_1$ and A$_2$ pseudoatoms shows two peaks at $E_{\rm F}$ and around $-$1.1 eV, closely resembling those of the La LDOS. Thus, these LDOS patterns together with the orbital projected band structure (Fig. 3a) manifest that the electronic states across $E_{\rm F}$ show a hybridization of the La 5$d$ and A$_1$/A$_2$ anionic states.

Figure 4a shows the calculated total charge density of La$_0$H$_{10}$. It is seen that H atoms in the H$_{32}$ cage are bonded to each other with covalent bonds. Here, each H$-$H bond has a saddle point of charge density at its midpoint, similar to the C$-$C covalent bond in diamond~\cite{diamond}. The charge densities at the midpoints of the H$_{1}-$H$_1$ and H$_{1}-$H$_2$ bonds are 0.66 and 0.74 $e$/{\AA}$^3$, respectively. It is, however, noticeable that the calculated phonon spectrum of La$_0$H$_{10}$ exhibits negative frequencies in the whole Brillouin zone (see Fig. 5a), indicating that the H$_{32}$ cages without La atoms are dynamically unstable. Meanwhile, LaH$_{10}$ is thermodynamically stable at 300 GPa without any negative-frequency phonon mode (see Fig. 5b). It is thus likely that the stable H$_{32}$ cages in LaH$_{10}$ can be formed through capturing the loosely bound anionic electrons in the interstitial regions of LaH$_0$.

\begin{figure}[htb]
\centering{ \includegraphics[width=8.0cm]{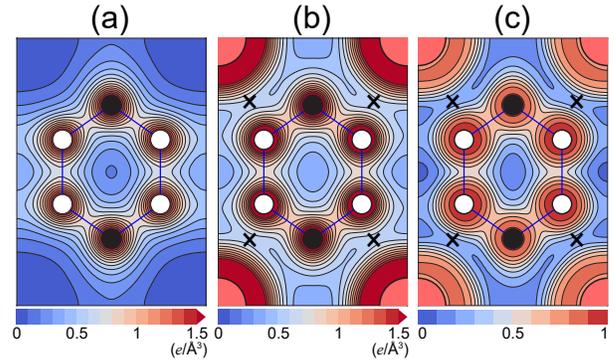} }
\caption{Calculated total charge densities of (a) La$_0$H$_{10}$ and (b) LaH$_{10}$ with the contour spacing of 0.1 $e$/{\AA}$^3$. In (b), the charge connection between La and H$_1$ atoms around the point marked ``${\times}$" represents the covalent character of the La$-$H$_1$ bond. The ELF of LaH$_{10}$ is also given in (c), with the contour spacing of 0.1.}
\end{figure}

In Fig. 4b, we display the total charge density of LaH$_{10}$, where the charge densities at the midpoints of the H$_{1}-$H$_1$ and H$_{1}-$H$_2$ bonds are 0.74 and 0.91 $e$/{\AA}$^3$, respectively. These values are larger than the corresponding ones (0.66 and 0.74 $e$/{\AA}$^3$) in La$_0$H$_{10}$ due to the addition of anionic electrons to the H$_{32}$ cages. Interestingly, we note that the polar H$_{1}-$H$_2$ covalent bond with a bond length $d_{\rm H_{1}-H_{2}}$ = 1.06 {\AA} is stronger than the nonpolar H$_{1}-$H$_1$ covalent bond with $d_{\rm H_{1}-H_{1}}$ = 1.15 {\AA}, possibly because H$_1$ atoms are connected to La atoms with some covalent bonding character, as discussed below. It is remarkable that donating electrons to the H$_{32}$ cages in LaH$_{10}$ can be supplied from the anionic electrons emerging in LaH$_0$. This feature of charge transfer driven by the electride property of the La framework is obviously distinct from the previous argument~\cite{rare-earth-hydride-PRL2017} that a charge transfer would occur from La to H atoms due to the higher electronegativity of H compared to that of La.

\begin{figure}[htb]
\centering{ \includegraphics[width=8.0cm]{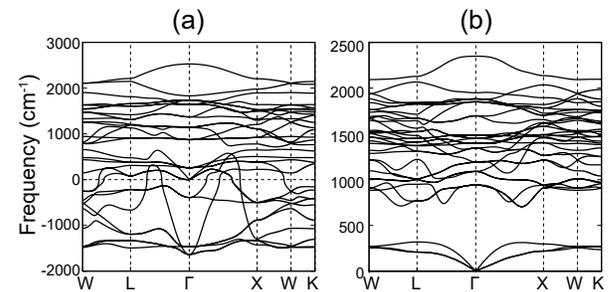} }
\caption{Calculated phonon spectra of (a) La$_0$H$_{10}$ and (b) LaH$_{10}$ at 300GPa.
}
\end{figure}

\begin{figure}[htb]
\centering{ \includegraphics[width=8.0cm]{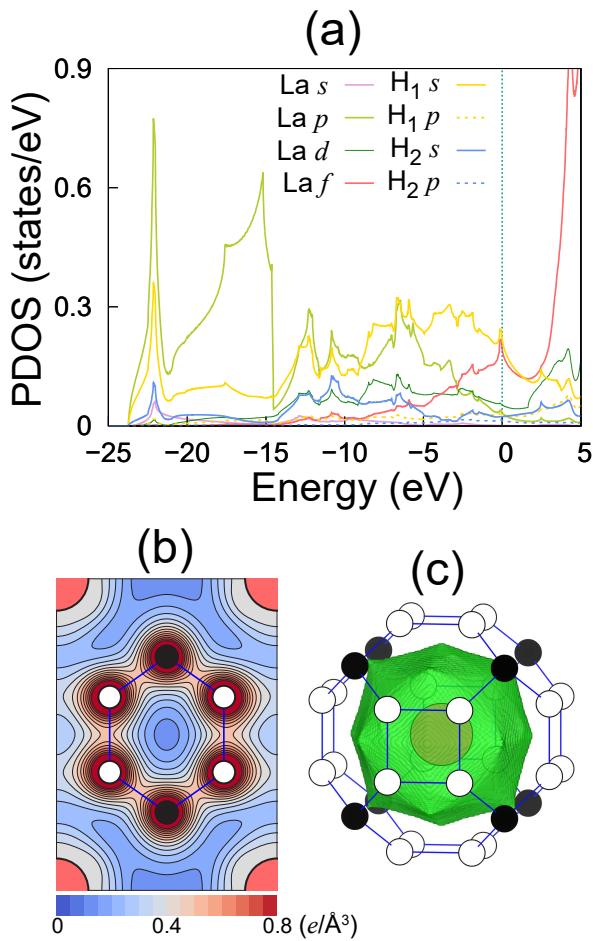} }
\caption{Calculated (a) PDOS of LaH$_{10}$ and (b) partial charge density within the energy range from $-$15 to $-$5 eV. In (b), the contour spacing is 0.05 $e$/{\AA}$^3$. Bader basin of La atom in LaH$_{10}$ is displayed in (c) with green color. }
\end{figure}

In order to examine the interaction between La atoms and H$_{32}$ cages, we calculate the partial DOS (PDOS) of LaH$_{10}$. From Fig. 6a, we find that the PDOS projected onto the La 5$p$ semicore states extends up to $E_{\rm F}$ with a significant hybridization with the H$_1$ 1$s$ state. Unlike the atomic-like La 5$p$ semicore states in LaH$_0$ (see Fig. S1 in the Supplementary Information), such delocalized semicore states in LaH$_{10}$ imply a strong interaction between La atoms and H$_{32}$ cages. Consequently, the electrical charges of La and H$_1$ atoms are connected with each other (see the point marked ``${\times}$" in Fig. 4b), indicating a covalent character between La and H$_1$ atoms. This weak covalent La-H$_1$ bonding character is also demonstrated by the ELF (Fig. 4c) of LaH$_{10}$ as well as the partial charge density (Fig. 6b) within the energy range from $-$15 to $-5$ eV where the La 5$p$ semicore and H$_1$ 1$s$ states are strongly hybridized with each other (see Fig. 6a). We note that the hybridization of La 4$f$ and H$_1$ 1$s$ states around $E_{\rm F}$ (see Fig. 6a) can also contribute to the La-H$_1$ covalent bonding character. Based on such covalent character of the La-H$_1$ bond as well as the addition of anionic electrons to the H$_{32}$ cages, we can say that the bonding nature between La atoms and H$_{32}$ cages is characterized as a mixture of ionic and covalent.

\begin{table}[hb]
\centering
\begin{tabular}{cccc}
\hline
            &  LaH$_0$ & La$_0$H$_{10}$ & LaH$_{10}$ \\ \hline
Q & $-$2.59 & 1.37       & $-$1.04   \\ \hline
\end{tabular}
\caption{ \textbf{Summed charge (Q) of Bader charge and nuclei charges outside the La basin in LaH$_0$, La$_0$H$_{10}$, and LaH$_{10}$} The La basin (see Fig. 6c) is equally employed for the Bader analyses of LaH$_0$, La$_0$H$_{10}$, and LaH$_{10}$.}
\label{tab:1}
\end{table}

Finally, we analyze the Bader charges~\cite{Bader} of LaH$_0$, La$_0$H$_{10}$, and LaH$_{10}$ to estimate the charge transfer between La atoms and H$_{32}$ cages. Figure 6c shows the Bader basin of La, obtained from a computation of the gradient of the density~\cite{Bader} in LaH$_{10}$. For the Bader analyses of LaH$_0$, La$_0$H$_{10}$, and LaH$_{10}$, we choose an identical La basin calculated from LaH$_{10}$. For LaH$_0$, the Bader charge outside the La basin can be considered as anionic electrons, while for LaH$_{10}$, it represents the electrons residing in the H$_{32}$ cage. For LaH$_0$, La$_0$H$_{10}$, and LaH$_{10}$, the summed charge (Q) of Bader charge and nuclei charges outside the La basin are listed in Table 1. We find that Q values for LaH$_0$, La$_0$H$_{10}$, and LaH$_{10}$ are $-$2.59$e$, 1.37$e$, and $-$1.04$e$, respectively. Therefore, LaH$_0$ behaves as an electride with a [La]$^{2.59+}$:2.59$e^{-}$ configuration. Here, the amount of anionic electrons is larger than that of [La]$^{1.83+}$:1.83$e^{-}$ estimated using the muffin-tin sphere around La atom (see Fig. 2b), because the La basin has a smaller volume than the La muffin-tin sphere. Meanwhile, the summation of Q values of LaH$_0$ and La$_0$H$_{10}$ is $-$1.22$e$, larger in magnitude than that ($-$1.04$e$) of LaH$_{10}$. This result indicates that the interaction between La atoms and H$_{32}$ cages in LaH$_{10}$ induces a small charge transfer of ${\Delta}$Q = 0.18$e$ from the H$_{32}$ cage to La atom. Such a reverse charge transfer is possibly related with the above-mentioned hybridization between the La 5$p$/4$f$ and H$_1$ 1$s$ states, which in turn leads to the charge connection between La and H$_1$ atoms around the ``${\times}$" point (see Fig. 4b).

\subsection{DISCUSSION}
Our first-principles DFT calculations for a room-temperature superconductor LaH$_{10}$ have shown that the bonding between La atoms and H$_{32}$ cages is characterized as a mixture of ionic and covalent. We revealed that the La framework has the anionic electrons at interstitial regions, which are easily captured to form the stable clathrate H$_{32}$ cages consisting of the nonpolar H$_{1}-$H$_1$ and polar H$_{1}-$H$_2$ covalent bonds. Specifically, the interaction between La atoms and H$_{32}$ cages involves a strong hybridization of the La 5$p$ semicore and H$_1$ 1$s$ states. It is thus demonstrated that anionic and semicore electrons are of vital importance to yield the intricate bonding nature in LaH$_{10}$. Our findings are rather generic and hence, they are anticipated to be more broadly applicable for understanding the bonding and electronic properties of other rare-earth hydrides with sodalitelike cage structures~\cite{rare-earth-hydride-PRL2017, rare-earth-hydride-PANS2017, YH10-Boeri, ExpCeH9-Nat.Commun2019T.Cui, ExpCeH9-Nat.Commun2019-J.F.Lin, ExpYH6-arXiv2019}.

\subsection{METHODS}
Our DFT calculations were performed using the Vienna {\it ab initio} simulation package with the projector-augmented wave method~\cite{vasp1,vasp2,paw}. Here, we treated La-5$s^2$5$p^6$5$d^1$6$s^2$ and H-1$s^1$ as valence electrons, with including 5$s^2$5$p^6$ semicore electrons for La. For the exchange-correlation energy, we employed the generalized-gradient approximation functional of Perdew-Burke-Ernzerhof~\cite{pbe}. A plane-wave basis was used with a kinetic energy cutoff of 500 eV. The ${\bf k}$-space integration was done with 24${\times}$24${\times}$24 $k$ points (in the Brillouin zone) for the structure optimization and 60${\times}$60${\times}$60 $k$ points for the DOS calculation. All atoms were allowed to relax along the calculated forces until all the residual force components were less than 0.005 eV/{\AA}. We calculated phonon frequencies with the 6${\times}$6${\times}$6 $q$ points using the QUANTUM ESPRESSO package~\cite{QE}.

\subsection{DATA AVAILABILITY}
The data that support the findings of this study are available from the corresponding author upon reasonable request.

\subsection{ACKNOWLEDGEMENTS}
This work was supported by the National Research Foundation of Korea (NRF) grant funded by the Korean Government (Grants No. 2019R1A2C1002975, No. 2016K1A4A3914691, and No. 2015M3D1A1070609). The calculations were performed by the KISTI Supercomputing Center through the Strategic Support Program (Program No. KSC-2019-CRE-0183) for the supercomputing application research.

\subsection{AUTHOR CONTRIBUTIONS}
J.H.C. designed and supervised the work. S. Y. and C. W. performed the theoretical calculations, and they contributed equally to this work. J.H.C., S. Y., C.W., and H. J. analyzed the data and wrote the manuscript. All authors contributed to the scientific discussion and manuscript revisions.

\subsection{ADDITIONAL INFORMATION}
\textbf{Supplementary information} is available for this paper at https://doi.org/10.1038/xxxxxxx.\\
\textbf{Competing financial interests} The authors declare no competing financial interests.

\clearpage
\onecolumngrid
\titleformat*{\section}{\LARGE\bfseries}

\renewcommand{\thefigure}{S\arabic{figure}}
\setcounter{figure}{0}

\vspace{1.2cm}

\section{Supplementary Information for "Stabilization mechanism of clathrate H cages in a room-temperature superconductor LaH$_{10}$"}
\begin{flushleft}
{\bf 1. PDOS of LaH$_0$}
\begin{figure}[ht]
\includegraphics[width=10cm]{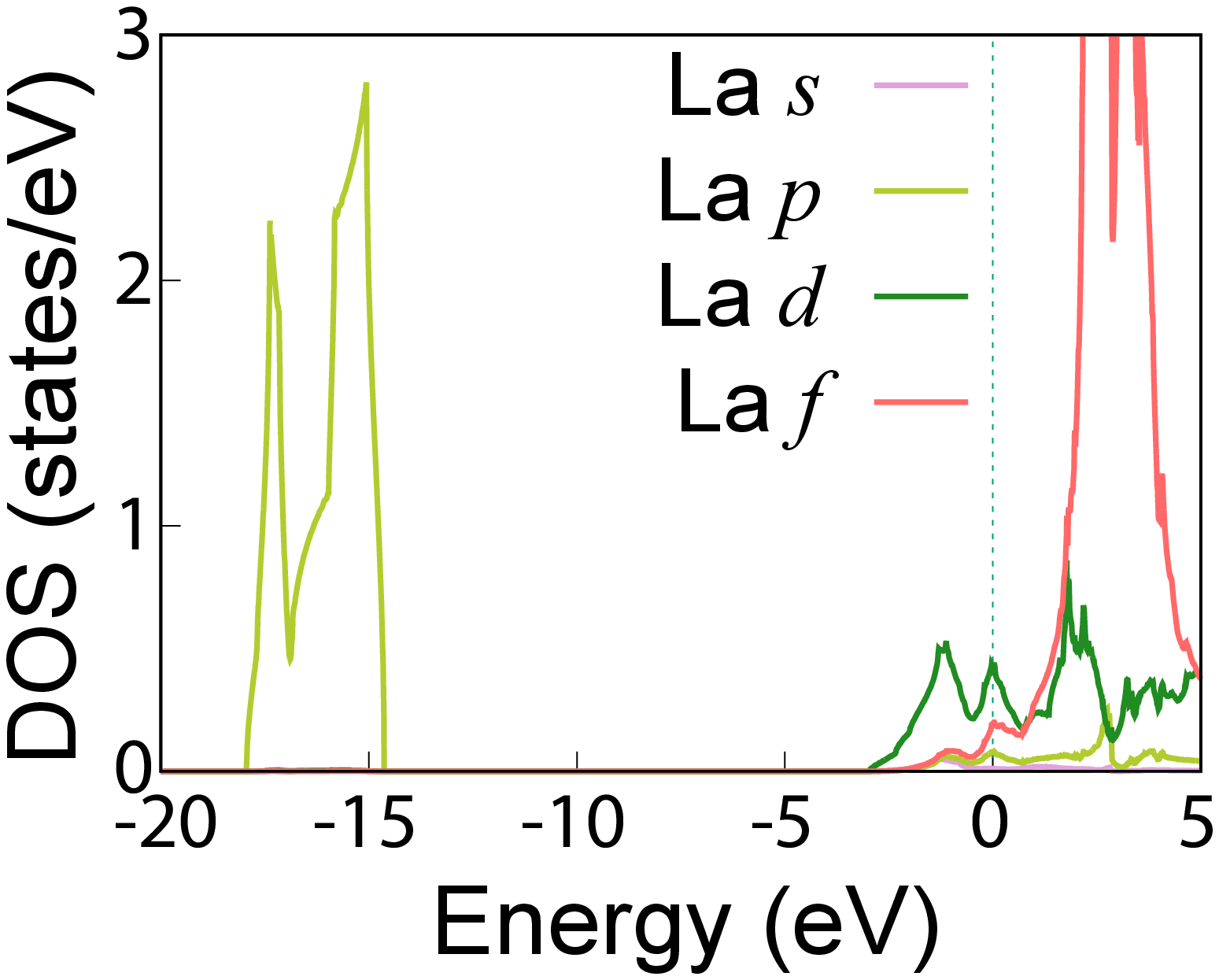}
\caption{ Calculated PDOS of LaH$_0$. The 5$s^2$ and 5$p^6$ semicore states locate at around $-$32 eV and $-$17 eV below $E_{\rm F}$.}
\end{figure}

\vspace{1.2cm}

{\bf 2. Two anions A$_1$ and A$_2$ in interstitial regions}
\begin{figure}[ht]
\includegraphics[width=12cm]{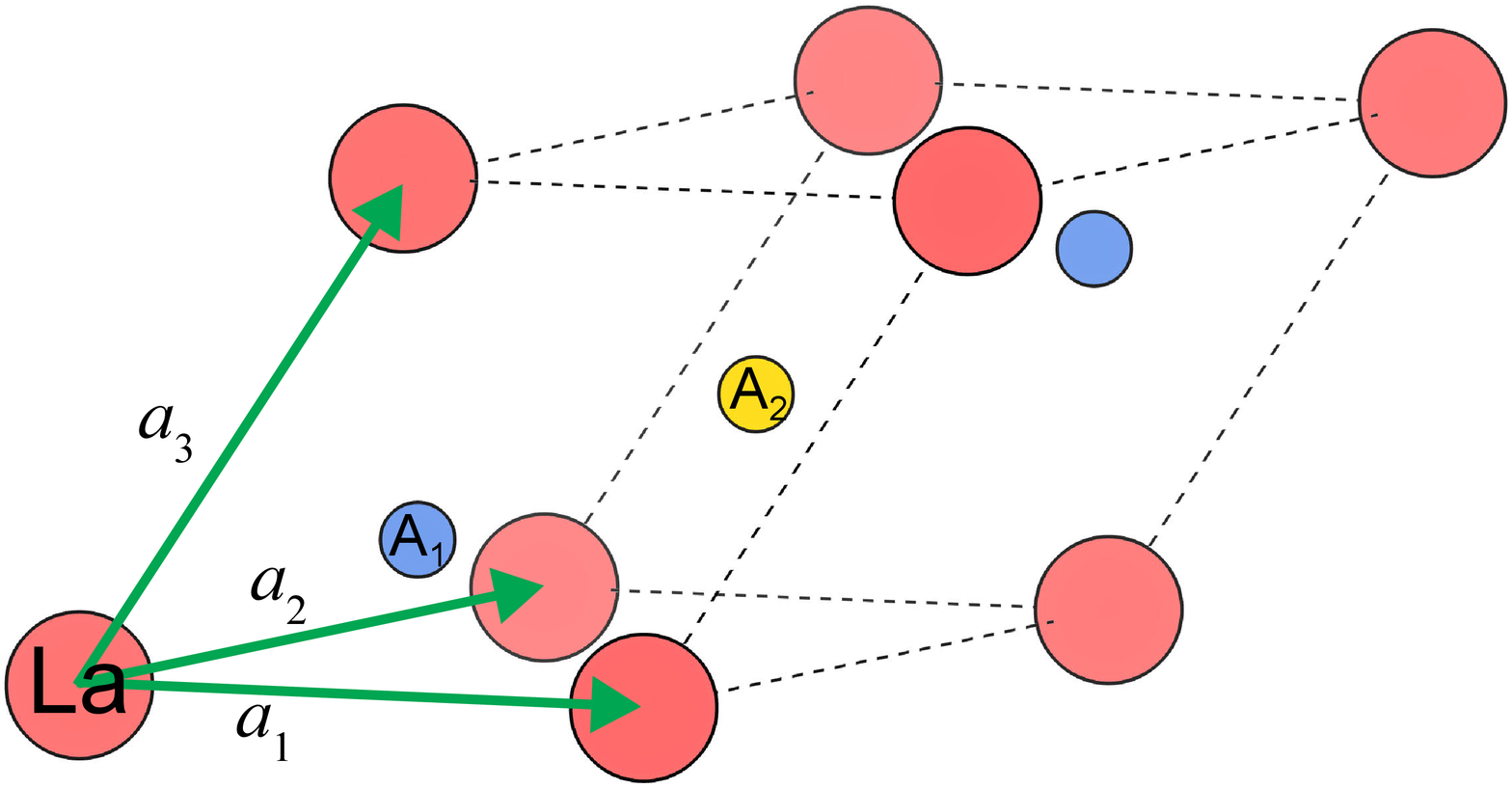}
\caption{ Location of the two anions A$_1$ and A$_2$ in LaH$_0$. There are two A$_1$ anions located at (0.25$a_1$, 0.25$a_2$, 0.25$a_3$) and (0.75$a_1$, 0.75$a_2$, 0.75$a_3$), respectively, and one A$_2$ anion at (0.5$a_1$, 0.5$a_2$, 0.5$a_3$).}
\end{figure}

\newpage

{\bf 3. Band projections onto the La, A$_1$, and A$_2$ (pseudo)atoms in LaH$_0$}
\begin{figure}[ht]
\includegraphics[width=\linewidth]{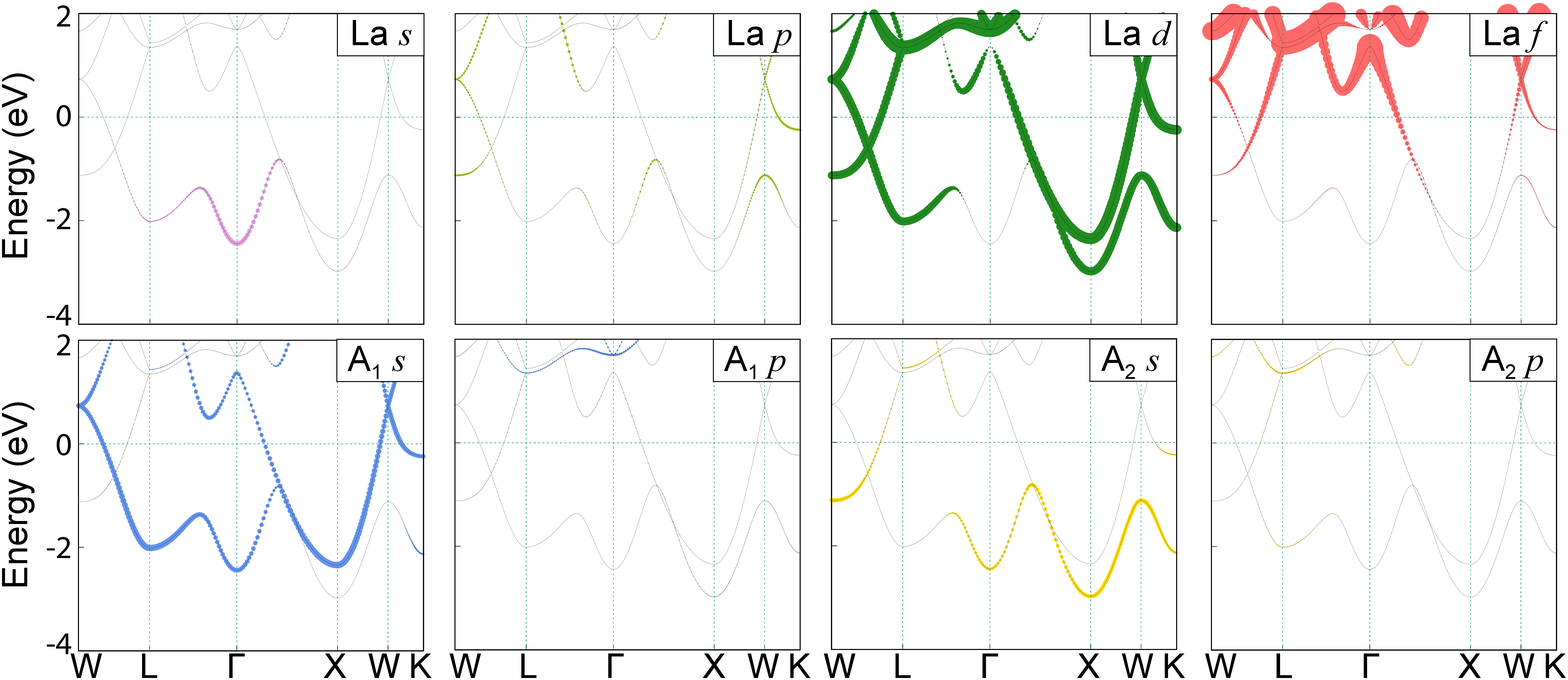}
\caption{ Calculated band structures of LaH0, projected onto the La $s$, La $p$, La $d$, La $f$, A$_1$ $s$, A$_1$ $p$, A$_2$ $s$ and A$_2$ $p$ orbitals. Here, the radii of circles are proportional to the weights of the corresponding orbitals.}
\end{figure}

\vspace{0.4cm}
\end{flushleft}

\end{document}